\documentclass[twocolumn,a4paper,amsmath,amssymb,showpacs,prb,superscriptaddress]{revtex4-2}

\usepackage{graphicx} 
\usepackage{color}
\usepackage{bm}
\usepackage{braket}
\usepackage[normalem]{ulem}

\begin{document}

\title{Optical anisotropy and electronic states in the pleochroic material Ca$_3$ReO$_5$Cl$_2$}
\author{Takumi Tsukihara}
\affiliation{Department of Materials Science, Graduate School of Engineering, Osaka Metropolitan University, Sakai, Osaka 599-8531, Japan}
\author{Ibuki Terada}
\affiliation{Department of Materials Science, Graduate School of Engineering, Osaka Metropolitan University, Sakai, Osaka 599-8531, Japan}
\author{Michi-To Suzuki}
\affiliation{Department of Materials Science, Graduate School of Engineering, Osaka Metropolitan University, Sakai, Osaka 599-8531, Japan}
\affiliation{Center for Spintronics Research Network, Graduate School of
Engineering Science, The University of Osaka, Toyonaka, Osaka 560-8531, Japan}

\begin{abstract}
Pleochroism is a type of optical anisotropy in which the apparent color of a material varies depending on the polarization and propagation direction of incident light. The oxychloride compound Ca$_3$ReO$_5$Cl$_2$ has recently attracted attention due to its pronounced pleochroism. The paramagnetic state of this compound, characterized by localized Re 5$d$ electrons, is challenging to describe within conventional first-principles methods.
In this study, we investigate the optical anisotropy in Ca$_3$ReO$_5$Cl$_2$ using first-principles calculations, with particular focus on the relationship between the optical spectra and electronic states.
We employ a ferromagnetically ordered state to effectively capture the localized character of the Re 5$d$ electrons. The calculated dielectric function and absorption coefficient qualitatively reproduce the experimentally observed peak structures.
An orbital-resolved analysis indicates that the characteristic optical transitions associated with the pleochroism predominantly involve Re-$d$-dominated electronic states, highlighting the key role of the Re $d$ electrons in the pleochroic optical response of Ca$_3$ReO$_5$Cl$_2$.
\end{abstract}

\maketitle

\section{Introduction}
Optical anisotropy, i.e., the dependence of the refractive index and absorption coefficient on the propagation direction and polarization of light, plays a central role in the functionality of optical materials and devices. Pleochroism is a form of optical anisotropy in which a material exhibits different colors depending on the polarization and propagation direction of incident light. 
This phenomenon is well known in natural minerals such as tourmaline and topaz~\cite{Manning1968,Hurlbut1969}.

Ca$_3$ReO$_5$Cl$_2$~(CROC) is a $5d$ transition-metal compound synthesized in $2017$ that exhibits remarkably strong pleochroism compared with natural minerals~\cite{Hirai2017}.
When the polarization of incident light is aligned along the $a$, $b$, and $c$ crystallographic axes, the compound appears green, red, and yellow, respectively. This pronounced optical anisotropy is believed to originate from both $d$–$d$ transitions of the Re ions and charge-transfer transitions between the Re ions and the surrounding ligands.
In addition to its optical properties, CROC has attracted interest as a frustrated quantum magnet composed of Re 5$d^1$ moments on an anisotropic triangular lattice~\cite{Hirai2019,Nawa2020,Zvyagin2022}.
Density functional theory (DFT) calculations without magnetic ordering predict a metallic electronic structure~\cite{Hirai2017}, whereas experimental studies suggest that CROC is a Mott insulator hosting localized spin-$1/2$ moments~\cite{Hirai2017}. 
At low temperatures, CROC undergoes a complex magnetic transition characterized by the propagation vector $\bm{q}=(0, 0.465, 0)$. In contrast, optical absorption measurements have mainly been performed at room temperature, where the system is in a paramagnetic state.
Accurately describing such a paramagnetic state with localized $d$ electrons within first-principles approaches remains challenging, making the microscopic origin of the pronounced pleochroism in CROC still unclear~\cite{Abrikosov2016}.

In this study, we investigate the microscopic origin of the optical anisotropy in CROC by analyzing the relationship between its dielectric response and electronic transitions involving Re $d$ orbitals.
To capture the localized nature of the Re $d$ electrons in the paramagnetic phase, we employ spin-polarized DFT calculations.
Based on the obtained electronic structure, we calculate the dielectric function and absorption coefficient, and show that the experimentally observed peak structures are qualitatively reproduced.
Our analysis reveals that these optical anisotropy originates from transitions between specific Re-$d$-dominated electronic states, providing a microscopic understanding of the pronounced pleochroism in CROC and clarifying the role of magnetism in its optical response.

\section{Calculation Methods}
\begin{figure}
    \centering
    \includegraphics[width=1.0\linewidth]{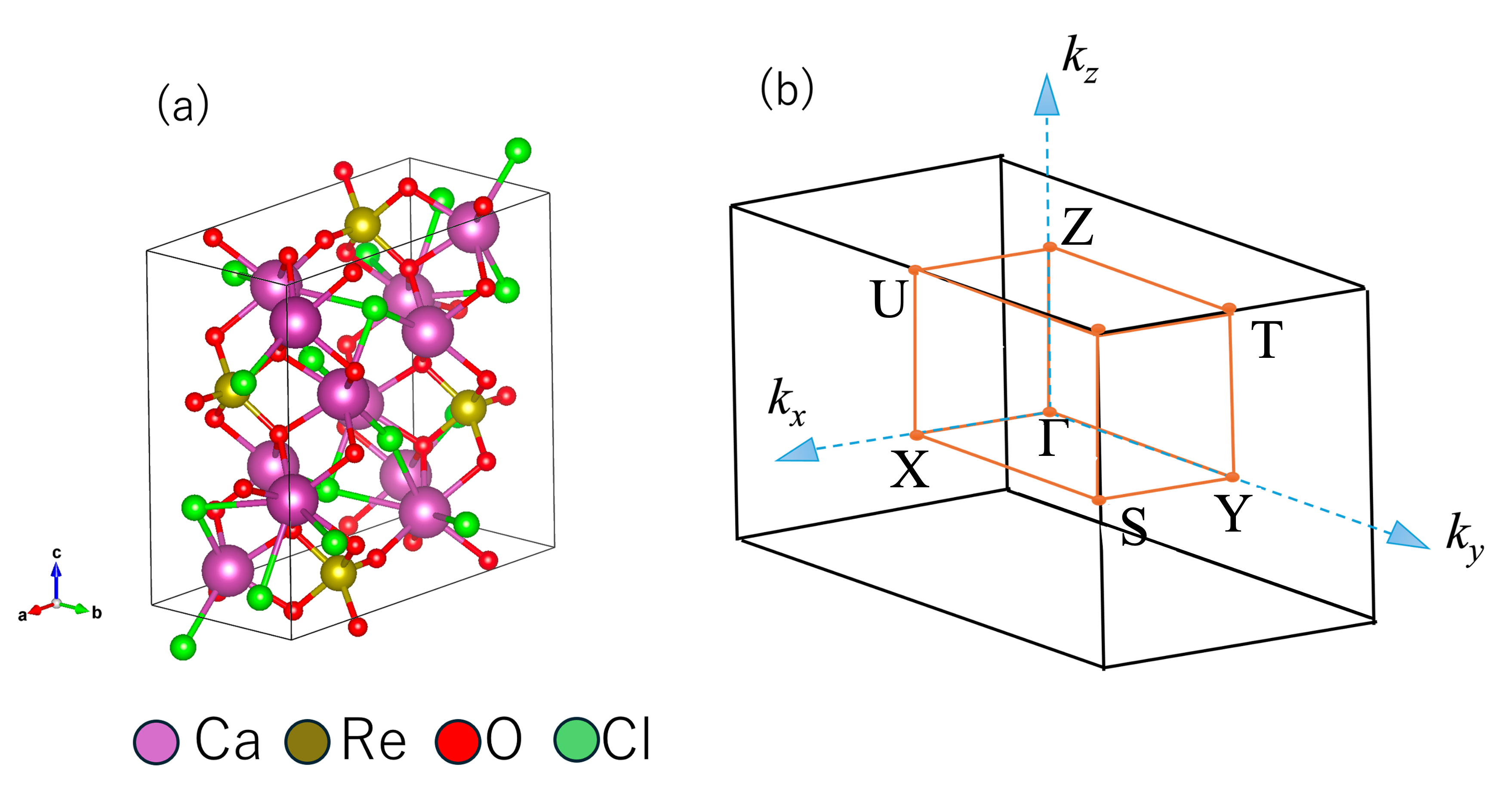}
    \caption{(a) Crystal structure and (b) Brillouin zone of CROC.}
    \label{fig:crystal}
\end{figure}

Figure~\ref{fig:crystal}(a) shows the crystal structure of CROC, which crystallizes in the orthorhombic crystal system with the space group $Pnma$. 
The experimental lattice constants are $a = 11.8997$~\text{\AA}, $b = 5.5661$~\text{\AA}, and $c = 11.1212$~\text{\AA}~\cite{Hirai2017}.
The unit cell contains 12 Ca, 4 Re, 20 O and 8 Cl atoms, giving 44 atoms per unit cell.
The experimentally determined lattice constants and Wyckoff positions reported in Ref.~\cite{Hirai2017} were used as the input crystal structure in the present calculations.

First-principles calculations were performed using the VASP code~\cite{Kresse1993_1, Kresse1993_2,Kresse1996CMS}. 
The projector augmented-wave (PAW) method and the Perdew–Burke–Ernzerhof (PBE) exchange–correlation functional were employed. 
Spin-orbit coupling was included in all calculations.
The plane-wave cutoff energy was set to $300~\mathrm{eV}$, and a Monkhorst-Pack $3 \times 6 \times 3$ $k$-point mesh was used to sample the first Brillouin zone shown in Fig.~\ref{fig:crystal}(b).
 
Experimentally, pleochroism in CROC is observed in the paramagnetic phase. 
A realistic first-principles description of such a state generally requires computationally demanding approaches, such as large supercells or disordered local moment methods, which are impractical for CROC because of its large unit cell.
To incorporate the localized nature of the Re $d$ electrons expected in the paramagnetic state, we therefore performed spin-polarized calculations assuming ferromagnetically aligned magnetic moments on the Re atoms.
As shown later, the calculated optical absorption coefficients qualitatively reproduce the experimentally observed pleochroic behavior.

To evaluate the optical properties of CROC, we calculated the frequency-dependent dielectric tensor $\varepsilon_{\mu\nu}(\omega)$ using the VASP code, which describes the response along the $\mu$ direction to an electric field applied along the $\nu$ direction. Here, $\mu,\nu=x,y,z$ correspond to the crystallographic $a$-, $b$- and $c$-axes, respectively. 
The imaginary part of the dielectric function, $\varepsilon''_{\mu\nu}(\omega)$, is given as follows:
\begin{align}
&\varepsilon''_{\mu\nu}(\omega)
=\frac{4\pi^2 e^2}{\Omega}
\lim_{\bm{q} \to 0} \frac{1}{q^2} 
\sum_{c,v,\bm{k}} 2 w_{\bm{k}} 
\delta(\varepsilon_{c\bm{k}} - \varepsilon_{v\bm{k}} - \omega)\\
&\quad\quad\quad\quad\quad
\times \Braket{u_{c,\bm{k} + \bm{e}_\mu q}|u_{v,\bm{k}}}
\Braket{u_{v,\bm{k}}|u_{c,\bm{k} + \bm{e}_\nu q}},
\label{formula:imaginary}
\end{align}
where $\ket{u_{n,\bm{k}}}$ represents the cell-periodic part of the $n$th Bloch wavefunction. The subscript $c~(v)$ denotes the conduction~(valence) band state. The term $w_{\bm{k}}$ denotes the weight function at $\bm{k}$ point.
The real part of the dielectric function is obtained from the Kramers-Kronig relation, 
\begin{align}
&\varepsilon'_{\mu\nu}(\omega) 
= 1 + \frac{2}{\pi} \, \mathcal{P} 
\int_{0}^{\infty} \frac{\omega_1 \, \varepsilon''_{\mu\nu}(\omega_1)}
{\omega_{1}^2 - \omega^2} \, d\omega_1,
\label{formula:real}
\end{align}
where $\mathcal{P}$ denotes the Cauchy principal value~\cite{Wooten1972}.
In the space group $Pnma$, the off-diagonal components of the dielectric tensor vanish by symmetry.
Therefore, we consider only the diagonal elements of dielectric function, $\varepsilon_{\mu}$, where $\varepsilon_{\mu\nu}=\varepsilon_{\mu}\delta_{\mu\nu}$.
The absorption coefficient $\alpha_\mu(\omega)$ is given by
\begin{equation}
\begin{split}
\label{formula:absorb}
\alpha_\mu(\omega)&=\frac{2\omega}{c}\mathrm{Im}\sqrt{\varepsilon_\mu}\\
&=\frac{\sqrt{2}\omega}{c}\sqrt{\sqrt{\varepsilon'^{~2}_{\mu}(\omega)+\varepsilon''^{~2}_{\mu}(\omega)}-\varepsilon'_{\mu}(\omega)},
\end{split}
\end{equation}
where $c$ is the speed of light. 

\section{results}
\begin{figure}
    \centering
    \includegraphics[width=1\linewidth]{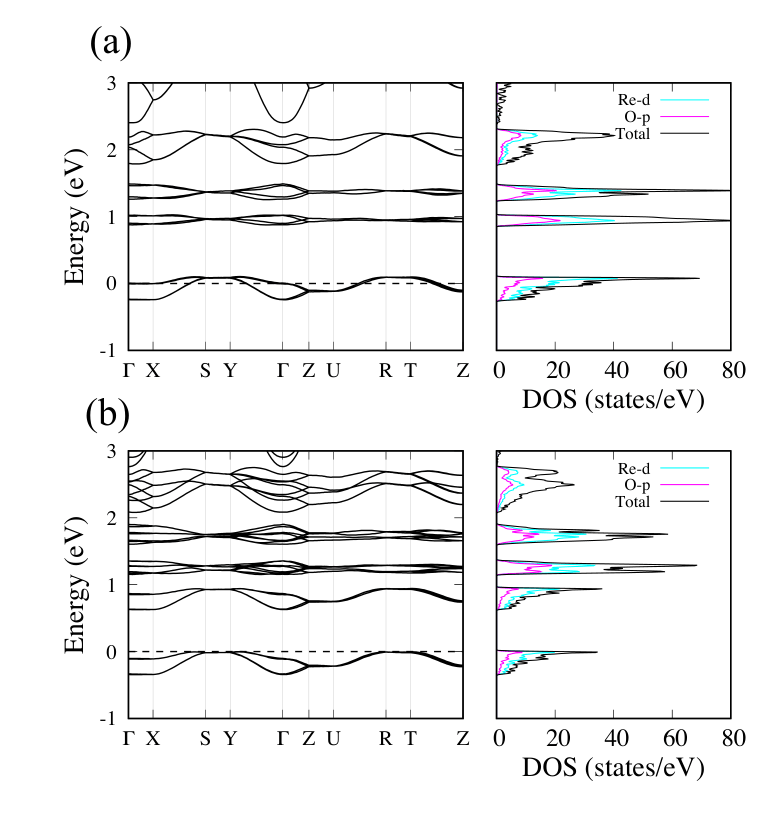}
    \caption{Band structurs (Left panels) and density of states (Right panels) of CROC calculated for (a) nonmagnetic and (b) spin-polarized calculations.}

    \label{Fig:band_dos}
\end{figure}

We first discuss the effects of the magnetic moment of the Re atom on the electronic structure of CROC. 
Figure~\ref{Fig:band_dos} shows the band structures and the density of states (DOS) for the nonmagnetic and ferromagnetic states.
No significant differences are found among calculations with the magnetic moments aligned along the $x$-, $y$-, and $z$-axes. 
All Re atoms carry a magnetic moment of $0.728~\mu_{B}$.

In the nonmagnetic calculation, CROC exhibits a metallic electronic structure, as shown in Fig.~\ref{Fig:band_dos}(a), consistent with previous studies~\cite{Hirai2017}.
This metallic behavior arises from the spin degeneracy inherent in the nonmagnetic calculation, which produces degenerate electronic bands near the Fermi level and prevents the opening of an energy gap.
In contrast, spin polarization lifts the spin degeneracy of the energy bands near the Fermi level and opens an energy gap, leading to the insulating states shown in Fig.~\ref{Fig:band_dos}(b), consistent with experimental observations. The calculated band gap is $0.63~\mathrm{eV}$.
Hereafter, we focus on the ferromagnetic state and investigate the optical properties of CROC based on this insulating electronic structure.

\begin{figure}
    \centering
    \includegraphics[width=1\linewidth]{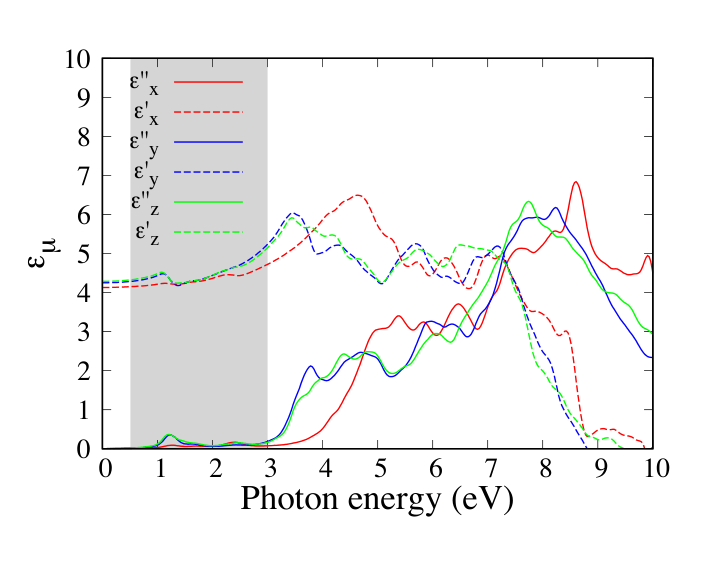}
    \caption{Real and imaginary parts of dielectric function in the photon energy range~$0-10~\mathrm{eV}$. The gray shaded region indicates the photon-energy range used in the experiment~\cite{Hirai2017}.}
    \label{fig:bigsmall}
\end{figure}

Figure~\ref{fig:bigsmall} shows the real and imaginary parts of the dielectric function $\varepsilon_{\mu}$, obtained from Eqs.~(\ref{formula:imaginary}) and (\ref{formula:real}), respectively. 
The dielectric function exhibits pronounced anisotropy over a wide photon-energy range.
The $y$- and $z$-components of the dielectric function exhibit broadly similar energy dependence over the entire energy range, whereas the $x$-component exhibits distinct behavior, indicating strong optical anisotropy in CROC.
The real part $\varepsilon'_{\mu}$ exhibits a peaks at approximately $3~\mathrm{eV}$ for the $y$- and $z$-components and at approximately $4.5~\mathrm{eV}$ for the $x$-component, followed by a sharp decrease above approximately $7.5~\mathrm{eV}$ for all components.
The imaginary part $\varepsilon''_{\mu}$ also shows strong anisotropy.
 The $y$- and $z$-components exhibit similar overall energy dependence, although noticeable differences emerge in the visible light region ($1.6–3.0~\mathrm{eV}$), as discussed later.
 These components increase significantly above $3~\mathrm{eV}$ and reaches a maxima near $7.5~\mathrm{eV}$. In contrast, the $x$-component begins to increase markedly just below $4~\mathrm{eV}$ and exhibits a peak at approximately $8.5~\mathrm{eV}$. 

\begin{figure}
    \centering
    \includegraphics[width=1.0\linewidth]{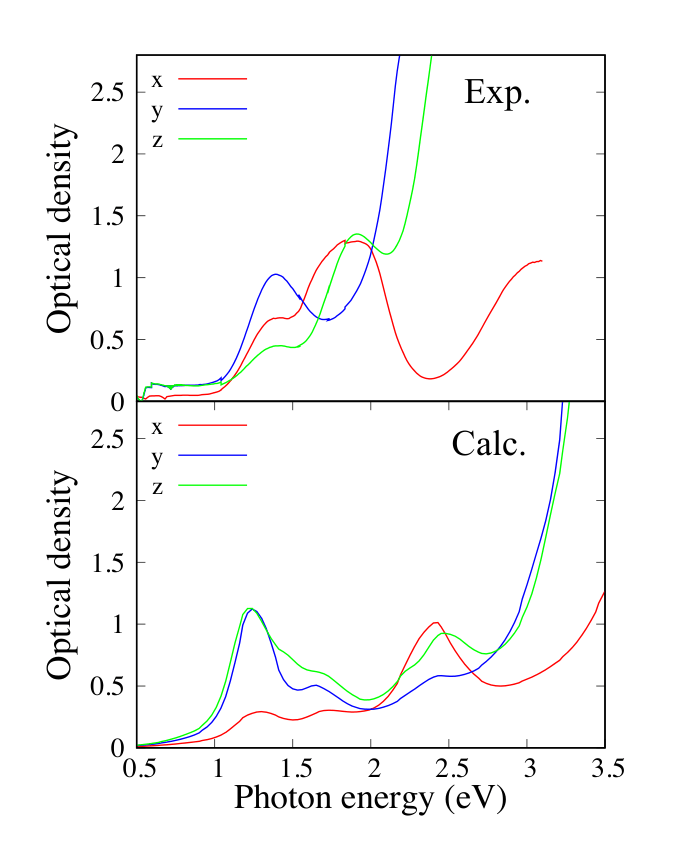}
    \caption{Optical density obtained from experiment~\cite{Hirai2017}~(Top) and numerical calculation (Bottom).}
    \label{fig:OD}
\end{figure}

Since pleochroism is observed in the visible light region of $1.6$–$3.0~\mathrm{eV}$ and experimental optical measurements are reported up to about $3~\mathrm{eV}$, we focus on this energy range.
To directly compare the calculated optical response with experiment, we evaluate the optical density (OD), defined as
\begin{align}
\label{formula:OD}
\mathrm{OD} &= \log_{10}\left(\frac{I_0}{I}\right) = \log_{10} \left( e^{\alpha d} \right) = \frac{\alpha d}{\ln 10},
\end{align}
where $I_0$ and $I$ are the incident and transmitted light intensities, respectively, $\alpha$ is the absorption coefficient, and $d$ denotes the sample thickness.
Figure~\ref{fig:OD} shows the experimental and calculated optical densities.
Panels (a) and (b) present the results in the photon-energy range from $0$ to $3~\mathrm{eV}$, where the sample thickness is set to $d = 2.4 \times 10^{-6}~\mathrm{m}$.
In this energy range, $\alpha_{\mu}$ is approximately proportional to the imaginary part of the dielectric function $\varepsilon''_{\mu}$ because $\varepsilon'_{\mu} \gg \varepsilon''_{\mu}$.

The calculated optical densities along the $x$-, $y$-, and $z$-axes exhibit a pronounced peak at approximately $1.3~\mathrm{eV}$. The peak energy is consistent with the experimental observations.
However, while the $y$- and $z$-components exhibit comparable peak intensities, the experimental data show a substantially weaker peak for the $z$-component.
This discrepancy may originate from the ferromagnetic approximation employed in the present calculations, which does not fully capture the magnetic disorder in the paramagnetic phase and may therefore affect the relative optical anisotropy in this energy range.

At higher photon energies, the calculated second peak and the rapid increase in the optical density are shifted by approximately $0.5~\mathrm{eV}$ toward higher energies relative to the experimental spectra.
Nevertheless, the overall anisotropic spectral features remain in qualitative agreement with experiment.
In particular, the $x$- and $z$-components exhibit pronounced peak structures, whereas the $y$-component shows no distinct peak.
This behavior is consistent with the experimental observations and indicates that the anisotropic optical response of CROC is well reproduced in the higher-energy visible-light region.

\begin{figure}
    \centering
    \includegraphics[width=1.0\linewidth]{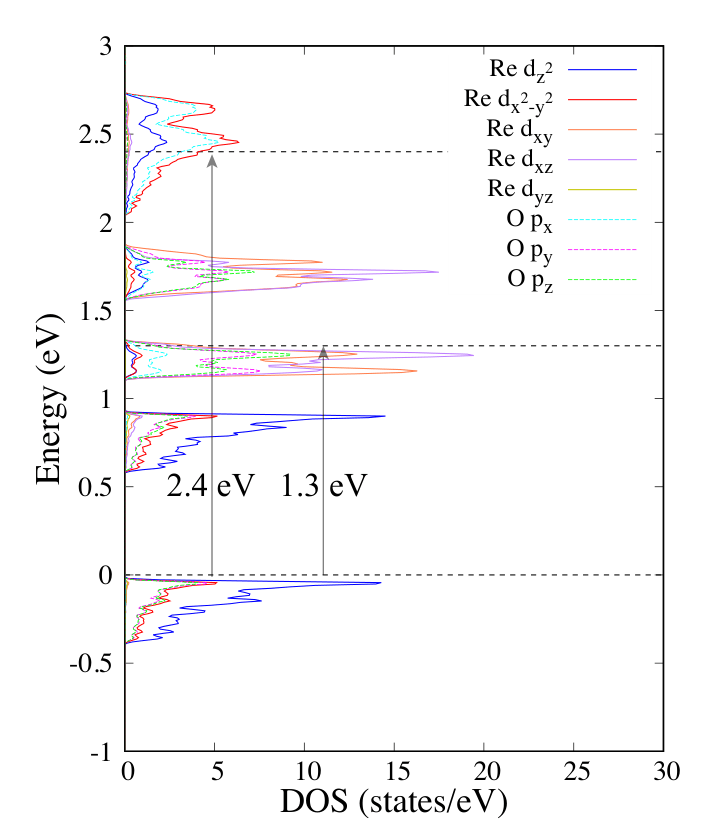}
    \caption{Partial density of states of CROC in magnetic states.}
    \label{fig:PDOS}
\end{figure}

To clarify the microscopic origin of the characteristic peaks in the optical density, we analyze the orbital character of the electronic states involved in the optical transitions.
As shown in the bottom panel of Fig.~\ref{fig:OD}, the calculated optical density exhibits two prominent peaks at photon energies of approximately $1.3~\mathrm{eV}$ and $2.4~\mathrm{eV}$.
To identify the electronic transitions responsible for these features, we examine the density of states projected onto the Re $d$ orbitals and the O $p$ orbitals, as shown in Fig.~\ref{fig:PDOS}.
The arrows in Fig.~\ref{fig:PDOS} indicate unoccupied electronic states located approximately $1.3~\mathrm{eV}$ and $2.4~\mathrm{eV}$ above the valence-band maximum, corresponding to the resonant peaks observed in the optical density.

As shown in Fig.~\ref{fig:PDOS}, the valence-band states near the valence-band maximum are mainly composed of Re $d_{z^2}$, Re $d_{x^2-y^2}$ orbitals hybridized with O $p_y$ and O $p_{z}$ orbitals.
The unoccupied states located approximately $1.3~\mathrm{eV}$ above the valence-band maximum, corresponding to the first resonant peak, exhibit substantial Re $d_{xy}$ and $d_{xz}$ character, in contrast to the valence-band states, while retaining strong O $p_y$ and $p_z$ contributions.
In contrast, the unoccupied states located approximately $2.4~\mathrm{eV}$ above the valence-band maximum, corresponding to the second resonant peak, retain the Re d-orbitals hybridized primarily with O $p_x$ orbitals.
The difference in orbital character and Re–O $d$–$p$ hybridization between the initial and final states generates finite optical transition matrix elements, leading to optical absorption at photon energies of approximately $1.3~\mathrm{eV}$ and $2.4~\mathrm{eV}$.
These optical excitations can therefore be interpreted as Re $d$–$d$ transitions mediated by oxygen $p$ orbitals, consistent with previous analyses based on maximally localized Wannier orbitals~\cite{Hirai2017}.

\begin{figure}
    \centering
    \includegraphics[width=1.00\linewidth]{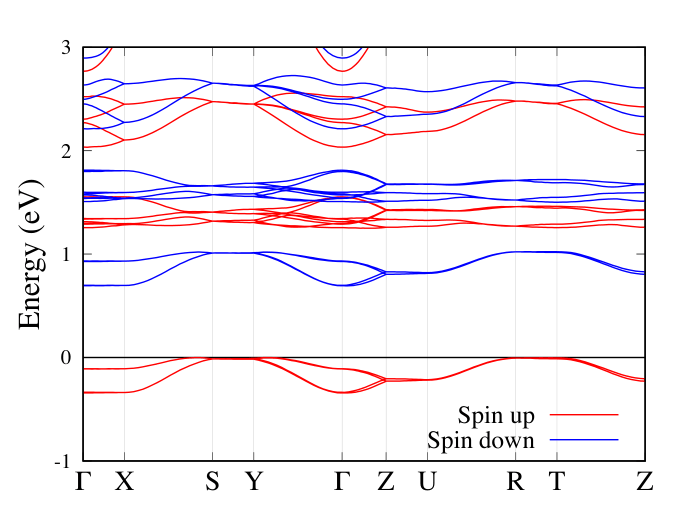}
    \caption{Spin-resolved band structure of CROC without spin–orbit coupling.}
    \label{fig:bandupdw}
\end{figure}

Optical transitions into the conduction-bands states located approximately $0.8~\mathrm{eV}$ and $1.7~\mathrm{eV}$ above the valence-band maximum are expected to be weak.
In the absence of spin–orbit coupling, the electronic bands are split into spin-up and spin-down channels, as shown in Fig.~\ref{fig:bandupdw}.
Since optical transitions approximately conserve spin, transitions between opposite spin states are strongly suppressed.
We confirmed that including spin–orbit coupling does not qualitatively alter the peak structure of the dielectric function.
This result indicates that the spin-resolved picture described above captures the essential physics of the optical transitions even in the presence of spin-orbit coupling.

\section{summary}
We investigated the microscopic origin of the pronounced pleochroism in Ca$_3$ReO$_5$Cl$_2$ by means of first-principles calculations.
Although pleochroism is observed experimentally in the paramagnetic phase, we adopted a ferromagnetically ordered configuration as an effective approximation to capture the localized nature of the Re $5d$ electrons.
Within this approximation, the spin splitting opens an energy gap and yields an insulating electronic structure consistent with experiment, in contrast to the metallic electronic structure obtained in nonmagnetic calculations.

Based on the resulting electronic structure, we calculated the dielectric function and optical absorption properties.
The calculated optical density reproduces the characteristic anisotropic features observed experimentally in the visible-light region, including resonant peaks at approximately $1.3~\mathrm{eV}$ and $2.4~\mathrm{eV}$.
Although quantitative discrepancies remain in the peak positions, the relative spectral intensities and polarization dependence are in qualitative agreement with experiment.

Orbital-resolved analysis revealed that the resonant optical features originate predominantly from transitions between Re $d$-dominated states mediated by anisotropic Re-O $d$-$p$ hybridization.
In addition, optical transitions between states with opposite spin character are strongly suppressed, indicating that spin selection plays an important role in the optical response.

These results demonstrate that the interplay between Re magnetic moments and anisotropic Re–O hybridization governs the pleochroic optical response of Ca$_3$ReO$_5$Cl$_2$.
Our findings provide a microscopic understanding of pleochroism in this $5d$ transition-metal compound and highlight the importance of magnetic degrees of freedom in shaping its optical anisotropy.

\begin{acknowledgements}
 We are grateful to K. Tanno for the technical supports.
 We also thank D. Hirai for providing experimental data.
 This research is supported by JSPS KAKENHI Grants Numbers JP23H01130, JP24K00581, JP24K00588, JP25K00947, JP25K21684, JP23K20824 and JP26K17065.
\end{acknowledgements}

\bibliography{reference}

\end{document}